\documentclass[11pt,a4paper]{article}
\usepackage{subcaption}
\usepackage{a4wide}
\usepackage{authblk}
\usepackage{graphicx}
\usepackage{xcolor}
\usepackage{amsmath,amsfonts,amssymb,amstext,graphicx}
\usepackage[colorlinks=true,  citecolor=blue, linkcolor=blue, urlcolor=black]{hyperref}
\usepackage[numbers,sort&compress]{natbib}
\usepackage{breakurl}
\usepackage{bm}

\usepackage{mathtools}
\usepackage{physics}

\usepackage{cancel}
\usepackage{empheq}
\usepackage{float} 
\usepackage{subcaption}
\usepackage{authblk}
\usepackage{comment}
\usepackage{blindtext}
\usepackage{cases}
\usepackage{ulem}

\usepackage[bottom]{footmisc}
\usepackage{lineno}

\usepackage{tikz}
\usetikzlibrary{arrows, positioning}

\def\k#1{\mathinner{|{#1}\rangle}}
\def\bra#1{\mathinner{\langle{#1}|}}
\def\ket#1{\mathinner{|{#1}\rangle}}

\begin{document}

\title{Spectral and Impedance Matching for Multiexcitation Charging of an Interacting Quantum Battery via an Anharmonic Mediator}

\author[1]{Sujay Mondal
\thanks{sujay.mondal.phy23@gm.rkmvu.ac.in}}

\author[2,3]{Anumita Mukhopadhyay
\thanks{anumitamukherjee455@gmail.com}}

\author[1]{Siddhartha Dutta
\thanks{siddhartha.dutta.phy23@gm.rkmvu.ac.in}}

\author[1]{Abhijit Bandyopadhyay
\thanks{abhijit.phy@gm.rkmvu.ac.in}}

\affil[1]{Department of Physics, Ramakrishna Mission Vivekananda
Educational and Research Institute,
Belur Math, Howrah 711202, West Bengal, India}

\affil[2]{Center for Quantum Engineering, Research and Education (CQuERE),
TCG CREST, Salt Lake, Sector 5, Kolkata 700091, India}

\affil[3]{Academy of Scientific and Innovative Research (AcSIR),
Ghaziabad 201002, India}

\date{\today}

\maketitle

\begin{abstract}
We investigate a finite-time quantum-battery  
charging architecture in
which a driven two-level working system transfers nonequilibrium resources
to an interacting two-qubit battery 
through a weakly anharmonic
three-level mediator. Finite-time driving 
generates coherence and
ergotropy, while excitation-number 
conservation organizes the transfer
into distinct dynamical sectors. 
The mediator then provides spectrally
selective charging channels: 
its lower transition enables nearly complete
single-excitation transfer with 
negligible residual mediator energy,
whereas initial mediator loading 
activates a cross-matched resonant
two-excitation cascade toward the 
doubly excited battery state.  In the latter sector, the mediator anharmonicity compensates the interaction-induced nonequidistance of the battery spectrum, enabling simultaneous resonance of the two transfer channels. At exact double resonance, unequal collective matrix elements impose the analytical bound $P_{\rm max}=8/9$; unit transfer is recovered under the impedance-matching condition $|G_1|=|G_2|$, as confirmed numerically.
  Detuning scans quantify the spectral tolerance, and the
full mixed-state dynamics reveals distinct energy- and ergotropy-transfer
profiles. Finite-time compression, 
thermal reset, and switching work close
the cycle-level energy balance to numerical 
precision. The resulting framework combines nonequilibrium resource generation, spectral matching, impedance matching, and coherent multilevel charging, and provides a basis for extending these principles to collective and many-body quantum batteries.
\end{abstract}

\section{Introduction}
\label{sec:intro}

Quantum batteries provide a framework for storing and transferring
energy in finite quantum systems, where the stored energy need not
coincide with the amount of useful work that can be extracted from the
state. The latter is quantified by the ergotropy
~\cite{Allahverdyan2004,AlickiFannes2013,Campaioli2024}, and the
distinction becomes particularly relevant when coherence,
nonequilibrium populations, and correlations contribute differently
to the battery state.  Collective light--matter and many-body interactions can enhance charging
power~\cite{Binder2015,Campaioli2017}, while theoretical and experimental
studies have demonstrated coherent quantum-energy storage in cavity,
solid-state, molecular, and superconducting platforms
~\cite{Ferraro2018,Gemme2024,Hu2026Charging}. More generally, recent work has emphasized that multilevel charging performance depends not only on spectral structure and transition connectivity, but also on the available control constraints and on how energy is partitioned among the charger, battery, and interaction channel~\cite{Murphy2026Energetic, Siddiqui2026Control}.\\

A complementary issue is how the energetic resource is transported
from a charger to the battery. Mediated charging has been studied with
bosonic and few-level intermediaries, including coherent quantum buses,
open ancillary systems, and sequential transfer protocols
~\cite{Andolina2018,Farina2019,Crescente2022,Crescente2023,ZhangJing2025}. More recently, structured photonic mediators have been shown to control
quantum-battery charging through bound-state formation, dark-state
protection, and topologically engineered energy and ergotropy transfer
~\cite{Lu2025TopologicalQB}.
In such architectures, the mediator is commonly regarded primarily as a transport element connecting an otherwise uncoupled charger and battery. For an interacting battery, however, the charging problem also involves the compatibility between the mediator spectrum and the distinct transition frequencies associated with different excitation sectors. This raises the more specific question of how the internal level structure of a finite-dimensional mediator can be used to selectively access the non-equidistant excitation gaps of an interacting battery. A multilevel mediator provides a natural setting for addressing this
problem, since its different transition frequencies can couple
selectively to different battery excitation sectors. Weakly anharmonic superconducting circuits, in particular, naturally provide
a spectrally resolved ladder with strong and tunable couplings
~\cite{Koch2007,Schreier2008,Blais2021CircuitQED}, and multilevel
quantum-battery protocols have already been implemented or proposed in
superconducting settings~\cite{Gemme2024,DouYang2023}. The mediator anharmonicity can therefore play a role beyond simple energy transport by providing an additional spectral degree of freedom with which the transition structure of the multilevel battery can be matched.\\

In this article, we investigate this spectral compatibility in a thermodynamically driven charging architecture. A two-level working system $S$, prepared in thermal equilibrium, is driven through a finite-time expansion that generates coherence and
ergotropy. At the end of this stroke, with the working-system Hamiltonian
held fixed, these resources are subsequently
transferred to an interacting two-qubit
battery $B$ through a weakly anharmonic 
three-level mediator $M$, with no
direct $S$--$B$ coupling. Conservation of the total excitation number
separates the transfer into distinct excitation sectors, allowing the role of the
mediator spectrum to be examined independently in each sector. We characterize the resonance structure and coherent transfer dynamics in two charging configurations, including their sensitivity to detuning
and the resulting redistribution of local
energy and ergotropy among the
working system, mediator, and battery. An unloaded mediator supports nearly complete single-excitation transfer through $S\rightarrow M\rightarrow B$ via its
ground-to-first-excited-state transition while leaving negligible residual resource in
the mediator. This configuration provides the
reference case in which only one mediator and one battery transition
need to be resonantly matched. Whereas, preparing the mediator
in its first excited state makes the transition to its second excited
state accessible and thereby opens a two-excitation pathway where both adjacent mediator transitions participate and must connect to the two distinct transition gaps of the interacting battery. The latter configuration therefore exposes a limitation that does not arise in the
single-excitation protocol: efficient multiexcitation transfer requires
simultaneous spectral compatibility across more than one transition.\\

The weak anharmonicity of the mediator provides the additional degree of
freedom required to address this constraint, that enables simultaneous cross matching of the two mediator and battery
transitions. Spectral resonance alone, however, does not completely determine the
transfer efficiency, as unequal collective matrix elements limit the
conditional end-to-end population transfer. This motivates a second level of optimization in which spectral matching
and coupling, or impedance matching are treated as distinct requirements
for efficient multiexcitation charging. \\

The resulting framework therefore separates the charging problem into
two complementary aspects: the spectral conditions that make the
relevant excitation pathways accessible and the dynamical matching
conditions that control population transfer along those pathways. Finally, finite-time
compression and rethermalization close the working-system cycle; when
the control work associated with the time-dependent transfer
couplings is included, the global energy balance is satisfied to
numerical precision.\\

The rest of the article is organized as follows.
Sec.~\ref{sec:model} describes finite-time resource generation in the
working system, Sec.~\ref{sec:SMB} develops the spectrally selective
mediator-assisted charging mechanism, and Sec.~\ref{sec:numerics}
presents the corresponding transfer and resource-redistribution
dynamics. Cycle closure through compression, thermal reset, and energy
accounting is discussed in Sec.~\ref{sec:transferandclosure}, followed
by the conclusions in Sec.~\ref{sec:conclusion}.

\section{Finite-time resource generation}
\label{sec:model}

\begin{figure*}[t]
    \centering
    \includegraphics[width=0.96\textwidth]{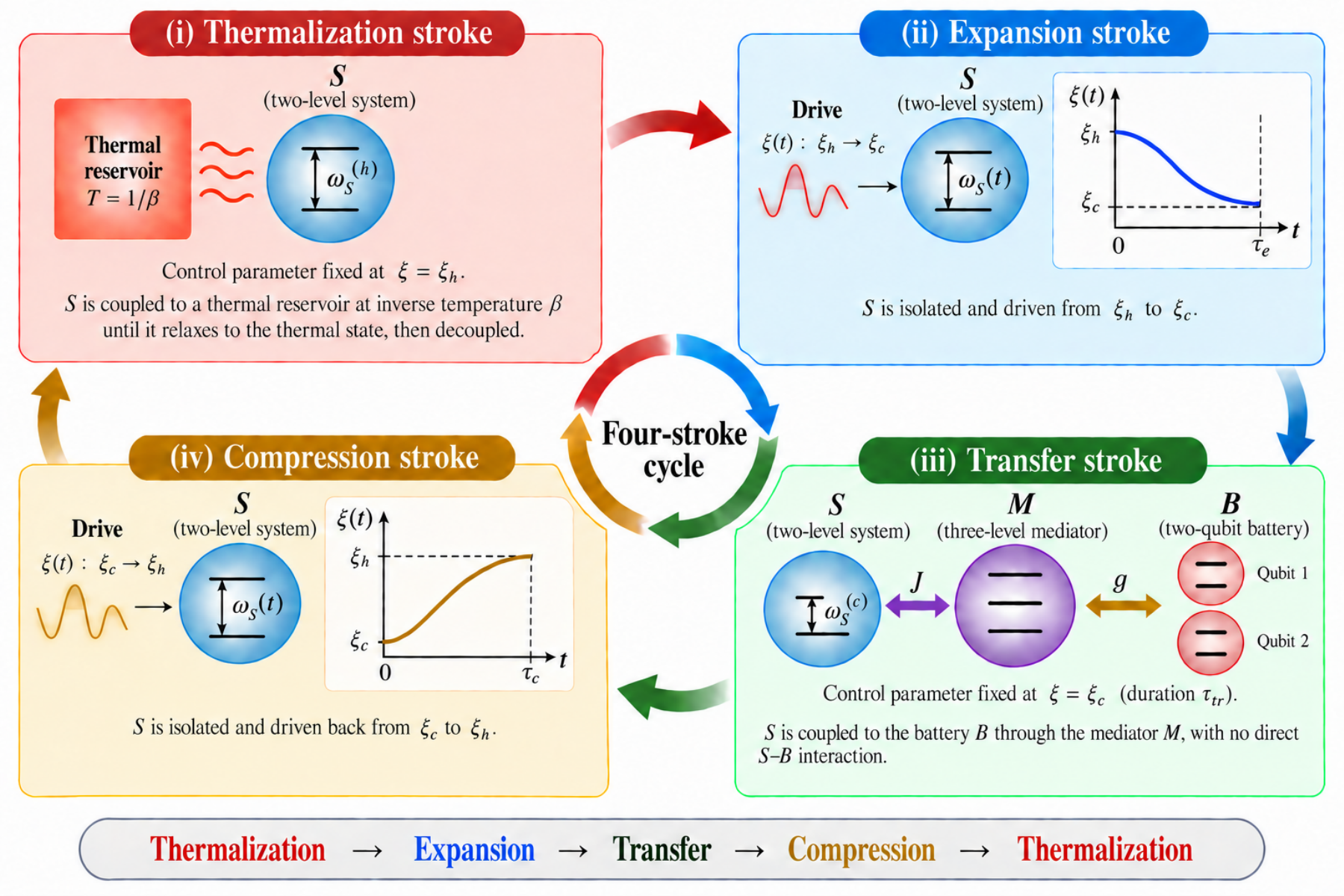}
    \caption{\textit{Mediator-assisted charging cycle.}
    (i) The working system $S$ is thermalized at fixed control
    $\xi_h$.
    (ii) An isolated finite-time expansion drives
    $\xi_h\rightarrow\xi_c$ over a duration $\tau_e$.
    (iii) At fixed $\xi_c$, the nonequilibrium resource generated in
    $S$ is coherently transferred to the interacting two-qubit battery
    $B$ through the three-level mediator $M$.
    (iv) The working system is isolated and compressed back to
    $\xi_h$, after which rethermalization restores its initial Gibbs
    state.}
    \label{fig:cycle}
\end{figure*}

The working system is a driven two-level system with Hamiltonian
\begin{eqnarray}
H_S(t)
=
\frac{1}{2}
\left[
\Delta_S\sigma_x^{(S)}
+
\xi(t)\sigma_z^{(S)}
\right]=
\frac{\omega_S(t)}{2}
\hat{\bm n}(t)\cdot\boldsymbol{\sigma}^{(S)},
\label{eq:HS_general}
\end{eqnarray}
where $\omega_S(t)=\sqrt{\Delta_S^2+\xi^2(t)}$
and
$\hat{\bm n}(t)=(\Delta_S,0,\xi(t))/\omega_S(t)$.
The transverse field $\Delta_S$ is fixed and the longitudinal
control $\xi(t)$ implements the expansion and compression strokes.
We use $\hbar=k_B=1$ and express all energies and frequencies in units
of a reference frequency $\omega_{\rm ref}$, set to unity; times are
therefore measured in units of $\omega_{\rm ref}^{-1}$.\\

At the beginning of the cycle, $\xi=\xi_h$ and
\begin{eqnarray}
H_S^{(h)}
&=&
\frac{1}{2}
\left(
\Delta_S\sigma_x^{(S)}
+
\xi_h\sigma_z^{(S)}
\right)
=
\frac{\omega_S^{(h)}}{2}
\hat{\bm n}_h\cdot\boldsymbol{\sigma}^{(S)},
\label{eq:HSh}
\end{eqnarray}
with
$\omega_S^{(h)}=\sqrt{\Delta_S^2+\xi_h^2}$ and
$\hat{\bm n}_h=(\Delta_S,0,\xi_h)/\omega_S^{(h)}$.
Introducing
$\cos\theta_h=\xi_h/\omega_S^{(h)}$ and
$\sin\theta_h=\Delta_S/\omega_S^{(h)}$, the eigenstates with
eigenenergies
$\varepsilon_\pm^{(h)}=\pm\omega_S^{(h)}/2$ are
\begin{eqnarray}
\k{\varepsilon_+^{(h)}}
&=&
\cos\!\left(\frac{\theta_h}{2}\right)\k{e_S}
+
\sin\!\left(\frac{\theta_h}{2}\right)\k{g_S},
\nonumber\\
\k{\varepsilon_-^{(h)}}
&=&
-\sin\!\left(\frac{\theta_h}{2}\right)\k{e_S}
+
\cos\!\left(\frac{\theta_h}{2}\right)\k{g_S},
\label{eq:HShes}
\end{eqnarray}
where $\{\k{e_S},\k{g_S}\}$ is the eigenbasis of
$\sigma_z^{(S)}$.
thermal contact with a reservoir 
at inverse temperature $\beta$
prepares $S$ in the Gibbs state of $H_S^{(h)}$,
\begin{eqnarray}
\rho_S^{(h),{\rm G}}
&=&
\frac{e^{-\beta H_S^{(h)}}}{Z_h},
\qquad
Z_h 
= {\rm Tr}
\left[
e^{-\beta H_S^{(h)}}
\right]
=
2\cosh\!\left(
\frac{\beta\omega_S^{(h)}}{2}
\right),
\label{eq:rhoGh1}
\end{eqnarray}
which may equivalently be expressed in Bloch form as
\begin{eqnarray}
\rho_S^{(h),{\rm G}}
&=&
\frac{1}{2}
\left[
\mathbb{I}
-
m_h\hat{\bm n}_h\cdot\boldsymbol{\sigma}^{(S)}
\right],
\qquad
m_h
=
\tanh\!\left(
\frac{\beta\omega_S^{(h)}}{2}
\right).
\label{eq:rhoGh}
\end{eqnarray}
Its populations in the energy basis are
$p_\pm^{(h)}=(1\mp m_h)/2$, and its energy is
\begin{eqnarray}
E_h
=
{\rm Tr}
\left[
\rho_S^{(h),{\rm G}}H_S^{(h)}
\right]
=
-\frac{m_h\omega_S^{(h)}}{2}.
\label{eq:Eh}
\end{eqnarray}
The Gibbs state is passive and therefore has vanishing ergotropy
~\cite{Allahverdyan2004}.\\

During the isolated expansion stroke, the control is changed from
$\xi_h$ to $\xi_c$ in a finite time $\tau_e$ according to
\begin{eqnarray}
\xi(t)
=
\xi_h
+
(\xi_c-\xi_h)
f\!\left(\frac{t}{\tau_e}\right),
\qquad
f(0)=0,
\quad
f(1)=1.
\label{eq:otto_protocol}
\end{eqnarray}
The final Hamiltonian is
\begin{eqnarray}
H_S^{(c)}
&=&
\frac{1}{2}
\left(
\Delta_S\sigma_x^{(S)}
+
\xi_c\sigma_z^{(S)}
\right)
=
\frac{\omega_S^{(c)}}{2}
\hat{\bm n}_c\cdot\boldsymbol{\sigma}^{(S)},
\label{eq:HSc}
\end{eqnarray}
where
$\omega_S^{(c)}=\sqrt{\Delta_S^2+\xi_c^2}$ and
$\hat{\bm n}_c=(\Delta_S,0,\xi_c)/\omega_S^{(c)}$.
The corresponding eigen states
$\k{\varepsilon_\pm^{(c)}}$ follow from
Eq.~\eqref{eq:HShes} under $h\rightarrow c$.
The expansion stroke takes the system
state from the initial Gibbs state $\rho_S^{(h),{\rm G}}$
to the state
\begin{eqnarray}
\rho_S^{(e)}
=
U_e\rho_S^{(h),{\rm G}}U_e^\dagger,
\qquad
U_e
=
{\cal T}
\exp\!\left[
-i\int_0^{\tau_e}dt\,H_S(t)
\right],
\label{eq:rho_after_exp}
\end{eqnarray}
with $H_S(t)$ given by Eq.\ \eqref{eq:HS_general}.
The instantaneous spectrum is
$\varepsilon_\pm(t)=\pm\omega_S(t)/2$, with eigenstates
\begin{eqnarray}
\k{\varepsilon_+(t)}
&=&
\cos\!\left[\frac{\theta(t)}{2}\right]\k{e_S}
+
\sin\!\left[\frac{\theta(t)}{2}\right]\k{g_S},
\nonumber\\
\k{\varepsilon_-(t)}
&=&
-\sin\!\left[\frac{\theta(t)}{2}\right]\k{e_S}
+
\cos\!\left[\frac{\theta(t)}{2}\right]\k{g_S},
\label{eq:HSinstant}
\end{eqnarray}
where the instantaneous mixing angle satisfies
$\cos\theta(t)=\xi(t)/\omega_S(t)$ and
$\sin\theta(t)=\Delta_S/\omega_S(t)$. 
The commutator
\begin{eqnarray}
\left[
H_S(t),H_S(t')
\right]
=
\frac{i\Delta_S}{2}
\left[
\xi(t)-\xi(t')
\right]
\sigma_y^{(S)},
\label{eq:Hcomm}
\end{eqnarray}
 is nonzero for $t\neq t'$ whenever $\Delta_S\neq0$ and
$\xi(t)\neq\xi(t')$. Consequently, a finite-rate variation of
$\xi(t)$ can induce nonadiabatic transitions between the instantaneous
energy branches, whereas such transitions are suppressed in the
adiabatic limit~\cite{Vitanov1996,Rezek2006,Camati2019}.
The local departure from
adiabatic following is  characterized  by
\begin{eqnarray}
{\cal A}(t)
=
\frac{|\dot{\theta}(t)|}{2\omega_S(t)}
=
\frac{|\Delta_S\dot{\xi}(t)|}
{2[\Delta_S^2+\xi^2(t)]^{3/2}},
\label{eq:adiabatic_parameter}
\end{eqnarray}
with ${\cal A}(t)\ll1$ in the adiabatic regime
~\cite{Landau1932,Zener1932,Vitanov1996,
Rezek2006,Camati2019}.\\

Since the expansion is unitary, the Bloch-vector magnitude is preserved
at $m_h$, and the final state can be written as
\begin{eqnarray}
\rho_S^{(e)}
=
\frac{1}{2}
\left[
\mathbb{I}
-
m_h\hat{\bm n}_e\cdot
\boldsymbol{\sigma}^{(S)}
\right],
\label{eq:rhoebloch}
\end{eqnarray}
where $\hat{\bm n}_e$ is the final Bloch-vector direction defined by $
U_e
\left(
\hat{\bm n}_h\cdot
\boldsymbol{\sigma}^{(S)}
\right)
U_e^\dagger
=
\hat{\bm n}_e\cdot
\boldsymbol{\sigma}^{(S)}$.
A convenient global measure of nonadiabaticity over the expansion stroke is provided by the nonadiabatic transition probability
\begin{eqnarray}
P_e
&=&
\left|
\langle\varepsilon_-^{(c)}
|U_e|
\varepsilon_+^{(h)}
\rangle
\right|^2
=
\frac{1-\hat{\bm n}_e\cdot\hat{\bm n}_c}{2}.
\label{eq:Pe_transition}
\end{eqnarray}
The misalignment between $\hat{\bm n}_e$ and the final Hamiltonian axis
$\hat{\bm n}_c$ also generates coherence 
in the final energy basis
$\ket{\varepsilon_\pm^{(c)}}$. Quantified by the
$\ell_1$ norm~\cite{Baumgratz2014,Streltsov2017,Lostaglio2015}, the
resulting coherence is 
\begin{eqnarray}
C_S^{(e)}
&=&
m_h
\sqrt{
1-
(\hat{\bm n}_e\cdot\hat{\bm n}_c)^2
}
=
2m_h\sqrt{P_e(1-P_e)}.
\label{eq:cohe}
\end{eqnarray}

The energies before and after the expansion are
\begin{eqnarray}
E_S^{(h)}
=
{\rm Tr}\!\left[
\rho_S^{(h),{\rm G}}H_S^{(h)}
\right]
=
-\frac{m_h\omega_S^{(h)}}{2},
\qquad
E_S^{(e)}
=
{\rm Tr}\!\left[
\rho_S^{(e)}H_S^{(c)}
\right]
=
-\frac{m_h\omega_S^{(c)}}{2}
\left(1-2P_e\right),
\label{eq:E_exp}
\end{eqnarray}
so the work performed on the isolated working system is
\begin{eqnarray}
W_e
=
E_S^{(e)}-E_S^{(h)}
&=&
\frac{m_h}{2}
\left[
\omega_S^{(h)}-\omega_S^{(c)}
\right]
+
m_h\omega_S^{(c)}P_e .
\label{eq:DeltaE_exp}
\end{eqnarray}
The first term is the adiabatic contribution, while the second is the
nonadiabatic excess energy. Since the eigenvalues of
$\rho_S^{(e)}$ are unchanged by the unitary stroke, its passive state
with respect to $H_S^{(c)}$ is
\begin{eqnarray}
\pi_S^{(e)}
=
\frac{1}{2}
\left[
\mathbb{I}
-
m_h\hat{\bm n}_c\cdot\boldsymbol{\sigma}^{(S)}
\right],
\label{eq:passive_expansion}
\end{eqnarray}
which gives the expansion-generated ergotropy
\begin{eqnarray}
{\cal W}_e
&=&
{\rm Tr}
\left[
\rho_S^{(e)}H_S^{(c)}
\right]
-
{\rm Tr}
\left[
\pi_S^{(e)}H_S^{(c)}
\right]
=
m_h\omega_S^{(c)}P_e.
\label{eq:We_Pe}
\end{eqnarray}
Thus, for the present two-level working system, the nonadiabatic excess
energy is entirely stored as ergotropy.\\

\begin{figure}[t]
\centering
\begin{subfigure}[b]{0.48\linewidth}
\centering
\includegraphics[width=\linewidth]{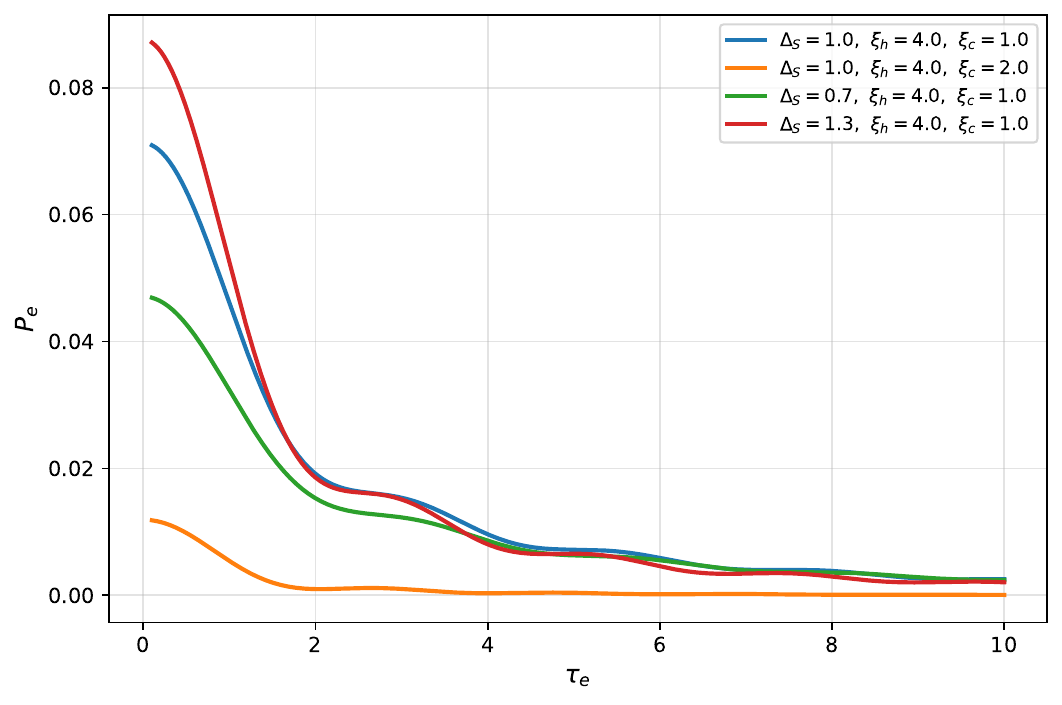}
\caption{}
\label{fig:panel_a}
\end{subfigure}
\hfill
\begin{subfigure}[b]{0.48\linewidth}
\centering
\includegraphics[width=\linewidth]{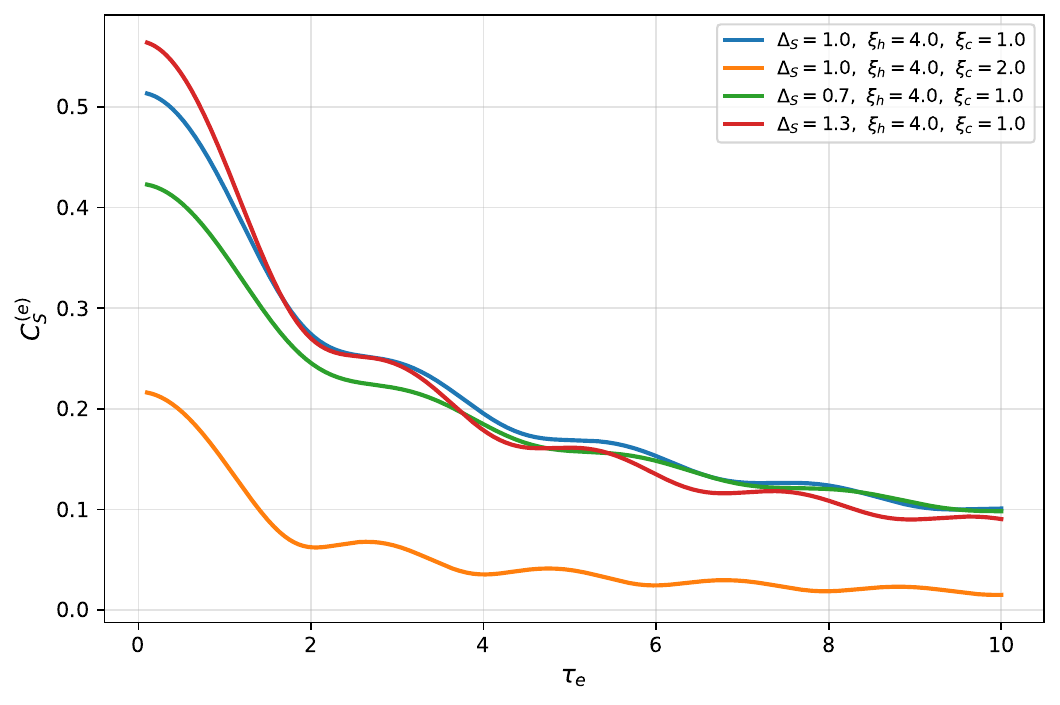}
\caption{}
\label{fig:panel_b}
\end{subfigure}

\vspace{0.3cm}

\begin{subfigure}[b]{0.48\linewidth}
\centering
\includegraphics[width=\linewidth]{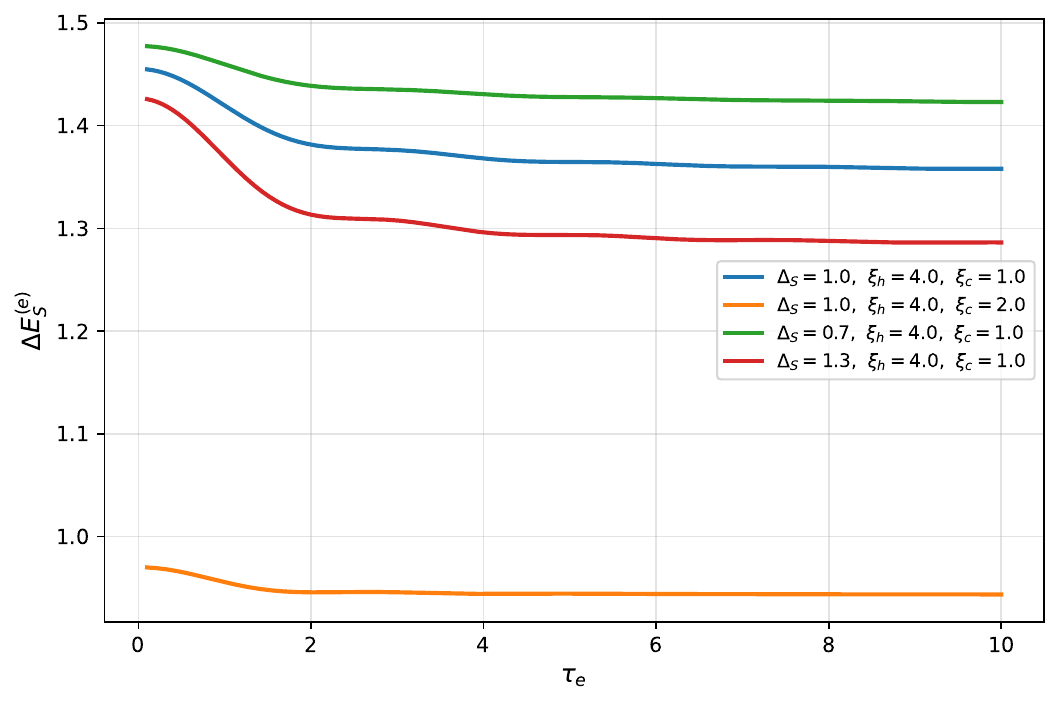}
\caption{}
\label{fig:panel_c}
\end{subfigure}
\hfill
\begin{subfigure}[b]{0.48\linewidth}
\centering
\includegraphics[width=\linewidth]{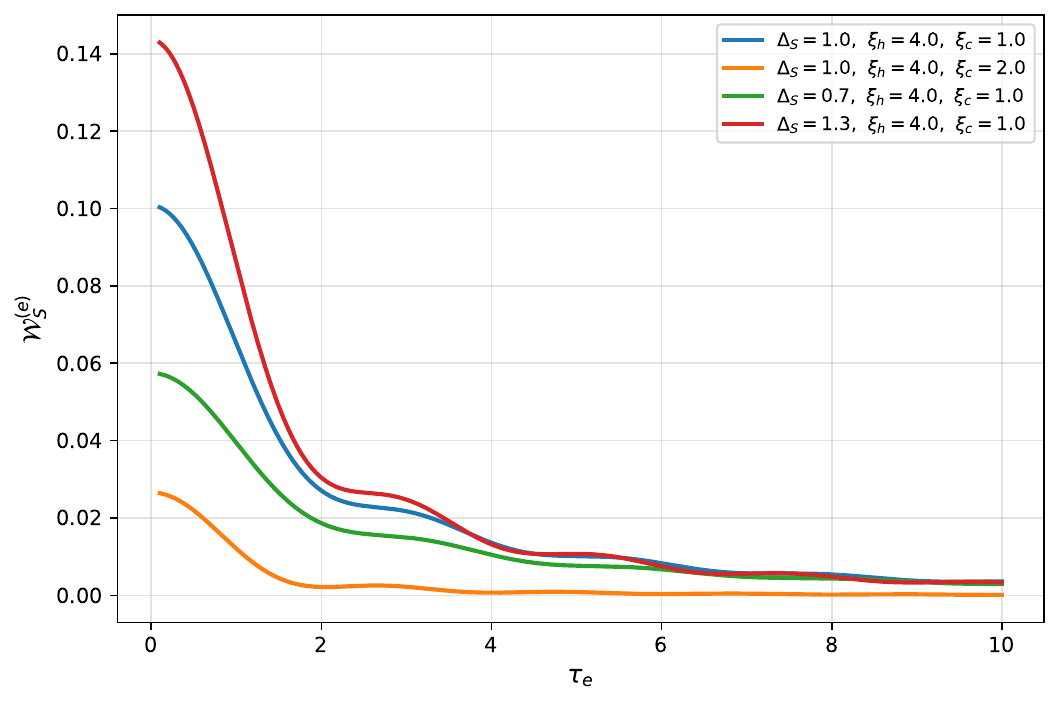}
\caption{}
\label{fig:panel_d}
\end{subfigure}
\caption{\textit{Finite-time resource generation during expansion.}
(a) Nonadiabatic transition probability $P_e$,
(b) coherence $C_S^{(e)}$ in the final energy basis,
(c) working-system energy change
$\Delta E_e=E_S^{(e)}-E_S^{(h)}$, and
(d) ergotropy ${\cal W}_e$ versus expansion duration $\tau_e$.
The initial Gibbs state has $\beta=5$ and $\xi_h=4$, and the
different curves correspond to the indicated values of
$(\Delta_S,\xi_c)$.}
\label{fig:expansion_resources}
\end{figure}
Fig.~\ref{fig:expansion_resources} shows the resource generated by a
linear protocol,
$f(t/\tau_e)=t/\tau_e$, over
$0.10\leq\tau_e\leq10.0$.
Since
$|\dot{\xi}|=|\xi_c-\xi_h|/\tau_e$, short expansion times correspond to
strong nonadiabatic driving, while increasing $\tau_e$ drives the dynamics
toward the adiabatic limit~\cite{Vitanov1996,Rezek2006,Camati2019}.
Accordingly,
short strokes produce finite $P_e$, coherence, and ergotropy, whereas
all three are suppressed as the dynamics approaches adiabatic
following. The total energy change instead approaches the finite
adiabatic contribution associated with the reduction of the
instantaneous gap. The expansion thus converts the initial thermal
population imbalance into a controllable nonequilibrium resource for
the subsequent charging stroke.

\section{Spectrally selective mediator-assisted transfer}
\label{sec:SMB}
At the end of the expansion, $\xi=\xi_c$ is held fixed while the
working system is coupled sequentially to the mediator and battery.
The transfer Hamiltonian is
\begin{eqnarray}
H_{\rm SMB}(t)
=
H_S^{(c)}
+
H_M
+
H_B
+
H_{SM}(t)
+
H_{MB}(t).
\label{eq:Hsmb}
\end{eqnarray}

The mediator $M$ is modeled as a weakly anharmonic three-level system
$\{\ket{0_M},\ket{1_M},\ket{2_M}\}$, as realized, for example, by the
lowest three levels of a transmon-type superconducting circuit
~\cite{Koch2007,Gambetta2011,Blais2021CircuitQED}.   
Setting the mediator ground-state energy to zero and introducing the
truncated ladder operator
$a_M =
\ket{0_M}\bra{1_M}
+
\sqrt{2}\ket{1_M}\bra{2_M}$, 
the mediator Hamiltonian is given by
\begin{eqnarray}
H_M
=
\omega_M\ket{1_M}\bra{1_M}
+
(2\omega_M+\alpha_M)\ket{2_M}\bra{2_M}
=
\omega_M a_M^\dagger a_M
+
\frac{\alpha_M}{2}
a_M^\dagger a_M
\left(a_M^\dagger a_M-1\right).
\label{eq:HM}
\end{eqnarray}
The frequencies for adjacent mediator transitions 
$\ket{0_M}\leftrightarrow\ket{1_M}$ and
$\ket{1_M}\leftrightarrow\ket{2_M}$ are
\begin{eqnarray}
\Omega_{M,1}
&=&
\omega_M,
\qquad
\Omega_{M,2}
=
\omega_M+\alpha_M ,
\label{eq:Mtranfreq}
\end{eqnarray}
respectively, with anharmonicity $\alpha_M<0$ for a transmon-like mediator. 
The interacting two-qubit battery is described by
the bare Hamiltonian
\begin{eqnarray}
H_B
&=&
\frac{\omega_B}{2}
\left(
\sigma_z^{(B_1)}
+
\sigma_z^{(B_2)}
\right)
+
\lambda_B
\left(
\sigma_+^{(B_1)}\sigma_-^{(B_2)}
+
\sigma_-^{(B_1)}\sigma_+^{(B_2)}
\right) 
+
\chi_B
\sigma_z^{(B_1)}
\sigma_z^{(B_2)} ,
\label{eq:HB}
\end{eqnarray}
where $\lambda_B$ and $\chi_B$ denote the exchange and Ising
couplings. Its eigenstates are
$\ket{00_B}$,
$\ket{11_B}$, and
\begin{eqnarray}
\ket{B_\pm}
=
\frac{1}{\sqrt{2}}
\left(
\ket{10_B}\pm\ket{01_B}
\right),
\end{eqnarray}
with eigenenergies
$E_{00}=-\omega_B+\chi_B$,
$E_{11}=\omega_B+\chi_B$, and
$E_\pm=-\chi_B\pm\lambda_B$.
The symmetric ladder
$\{\ket{00_B},\ket{B_+},\ket{11_B}\}$ therefore has adjacent
transition frequencies
\begin{eqnarray}
\Omega_{B,1}
&=&
\omega_B-2\chi_B+\lambda_B,
\qquad
\Omega_{B,2}
=
\omega_B+2\chi_B-\lambda_B .
\label{eq:Btranfreq}
\end{eqnarray}
The couplings $\lambda_B$ and $\chi_B$ therefore tune the battery
transition frequencies and provide the spectral matching required for
resonant excitation transfer~\cite{Andolina2019Ergotropy,
Mukherjee2016,Touil2021}.\\

The working system exchanges excitations with the mediator through~\cite{Jaynes1963,Blais2021CircuitQED}
\begin{eqnarray}
H_{SM}(t)
=
J(t)
\left[
\tau_+^{(S)}a_M
+
\tau_-^{(S)}a_M^\dagger
\right],
\label{eq:HSM}
\end{eqnarray}
where
$\tau_+^{(S)}
=
\ket{\varepsilon_+^{(c)}}\bra{\varepsilon_-^{(c)}}$,
$\tau_-^{(S)}=(\tau_+^{(S)})^\dagger$ and 
$J(t)$ denotes the controllable $S$--$M$ coupling.
The mediator couples collectively to the battery 
through the interaction 
~\cite{Tavis1968,Dicke1954,Ferraro2018,Farina2019}
\begin{eqnarray}
H_{MB}(t)
=
g(t)
\left[
a_M C_+^{(B)}
+
a_M^\dagger C_-^{(B)}
\right],
\label{eq:HMB}
\end{eqnarray}
where
$C_\pm^{(B)}=\sigma_\pm^{(B_1)}+\sigma_\pm^{(B_2)}$
are the collective battery ladder operators and $g(t)$ is the
controllable $M$--$B$ coupling. 
Exchange symmetry renders the antisymmetric 
state $\ket{B_-}$ dark,
$C_\pm^{(B)}\ket{B_-}=0$, while 
\begin{eqnarray}
C_+^{(B)}\ket{00_B}
=
\sqrt{2}\ket{B_+},
\qquad
C_+^{(B)}\ket{B_+}
=
\sqrt{2}\ket{11_B}.
\end{eqnarray}
The mediator therefore couples only to the bright 
symmetric ladder
$\ket{00_B}\leftrightarrow\ket{B_+}
\leftrightarrow\ket{11_B}$,
with a collective $\sqrt{2}$ 
enhancement of the allowed transition
matrix elements~\cite{Dicke1954,Tavis1968,Filipp2011}. \\

We use the nonoverlapping smooth pulses
~\cite{Bialczak2011,Sete2015,Li2021}
{\small
\begin{eqnarray}
J(t)
=
\begin{cases}
J_0
\sin^2\!\left(\dfrac{2\pi t}{\tau_{\rm tr}}\right),
&
0\leq t\leq\dfrac{\tau_{\rm tr}}{2},
\\[2mm]
0,
&
\dfrac{\tau_{\rm tr}}{2}
<t\leq\tau_{\rm tr},
\end{cases}
\qquad
g(t)
=
\begin{cases}
0,
&
0\leq t<\dfrac{\tau_{\rm tr}}{2},
\\[2mm]
g_0
\sin^2\!\left[
\dfrac{2\pi}{\tau_{\rm tr}}
\left(
t-\dfrac{\tau_{\rm tr}}{2}
\right)
\right],
&
\dfrac{\tau_{\rm tr}}{2}
\leq t\leq\tau_{\rm tr}.
\end{cases}
\label{eq:Jg}
\end{eqnarray}
}

The first half of the stroke loads the mediator from $S$, while the
second transfers the mediator excitation to $B$.\\

\begin{figure}[t]
    \centering
    \includegraphics[scale=0.22]{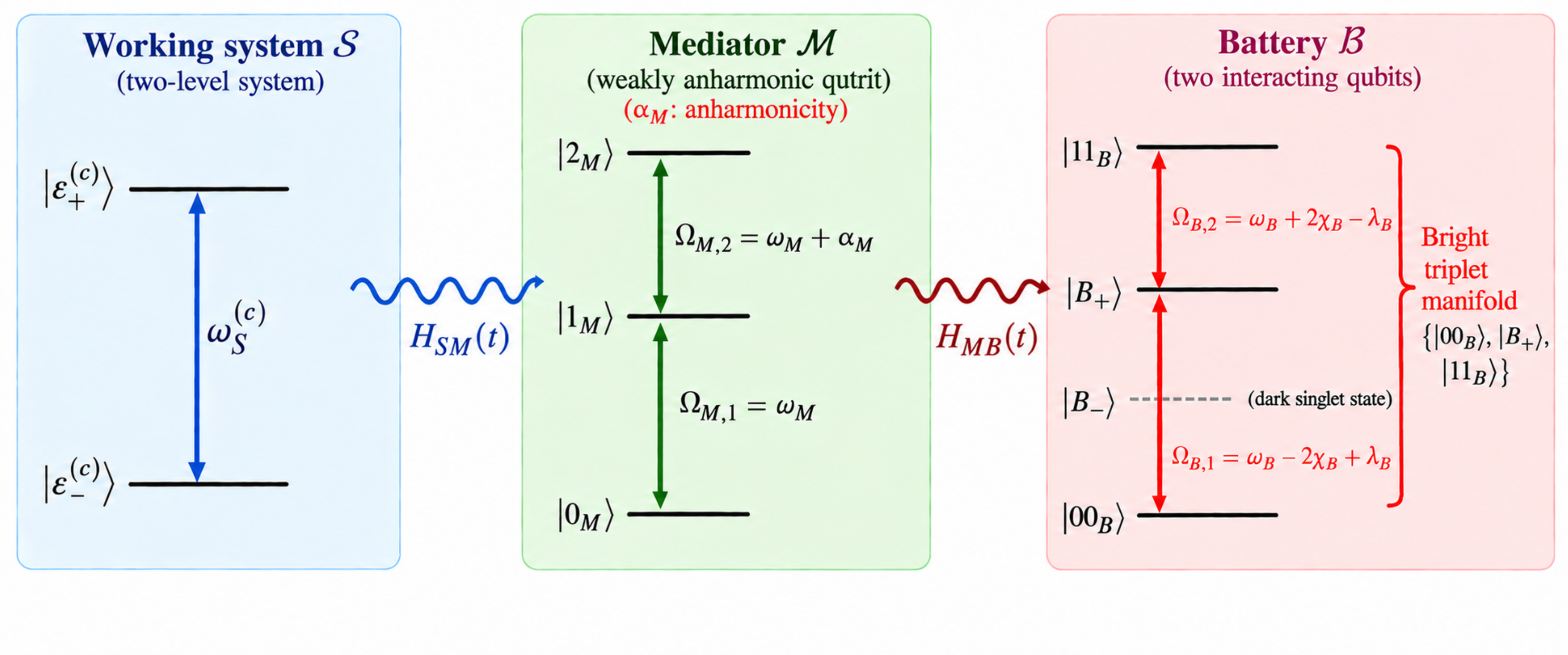}
    \caption{Energy-level structure during the transfer stroke.
    The couplings $J(t)$ and $g(t)$ mediate sequential
    $S$--$M$ and $M$--$B$ excitation exchange. The interacting battery
    is charged through the symmetric ladder
    $\ket{00_B}\leftrightarrow\ket{B_+}\leftrightarrow\ket{11_B}$,
    while $\ket{B_-}$ remains dark.}
    \label{fig:elevel}
\end{figure}

The underlying transfer structure follows directly from conservation
of the total excitation number,
\begin{eqnarray}
{\cal N}
=
\tau_+^{(S)}\tau_-^{(S)}
+
a_M^\dagger a_M
+
\sum_{j=1}^{2}
\sigma_+^{(B_j)}\sigma_-^{(B_j)},
\label{eq:Ntot}
\end{eqnarray}
for which
$[H_{\rm SMB}(t),{\cal N}]=0$.
The dynamics therefore decomposes into independent excitation sectors.
To characterize spectral matching within them, we define
\begin{eqnarray}
\delta_{SM}^{(r)}
&=&
\omega_S^{(c)}-\Omega_{M,r},
\qquad
\delta_{MB}^{(r,k)}
=
\Omega_{M,r}-\Omega_{B,k},
\qquad
r,k=1,2.
\label{eq:detunings}
\end{eqnarray}
The resulting level structure and
the corresponding resonant exchange 
pathways are illustrated in
Fig.~\ref{fig:elevel}. \\ \\
\subsection{Single-excitation charging protocol}
For an unloaded mediator and an initially empty battery, the active single-excitation pathway is
\begin{eqnarray}
\ket{\varepsilon_+^{(c)},0_M,00_B}
\longleftrightarrow
\ket{\varepsilon_-^{(c)},1_M,00_B}
\longleftrightarrow
\ket{\varepsilon_-^{(c)},0_M,B_+}.
\label{eq:path_single}
\end{eqnarray}
Its complete resonance condition is
\begin{eqnarray}
\omega_S^{(c)}
=
\Omega_{M,1}
=
\Omega_{B,1},
\qquad
\delta_{SM}^{(1)}
=
\delta_{MB}^{(1,1)}
=
0 .
\label{eq:single_double_res}
\end{eqnarray}
Only the lower mediator transition participates in this sector, so
$\alpha_M$ does not affect the corresponding resonant dynamics.\\
%


\subsection{Double-excitation charging protocol}
If instead the mediator starts in $\ket{1_M}$, the active component of
the working system enters the two-excitation sector,
\begin{eqnarray}
	\ket{\varepsilon_+^{(c)},1_M,00_B}
	\longleftrightarrow
	\ket{\varepsilon_-^{(c)},2_M,00_B}
	\longleftrightarrow
	\ket{\varepsilon_-^{(c)},1_M,B_+}
	\longleftrightarrow
	\ket{\varepsilon_-^{(c)},0_M,11_B}.
	\label{eq:path_double}
\end{eqnarray}
%
The two mediator transitions now connect the two different battery gaps. Resonance of
the complete two-excitation pathway requires
\begin{eqnarray}
	\delta_{SM}^{(2)}
	=
	\delta_{MB}^{(2,1)}
	=
	\delta_{MB}^{(1,2)}
	=
	0,
	\label{eq:double_MB_resonance}
\end{eqnarray}
which yields
\begin{eqnarray}
	\alpha_M
	=
	2(\lambda_B-2\chi_B),
	\qquad
	\omega_B
	=
	\omega_M+\frac{\alpha_M}{2}.
	\label{eq:double_res_parameters}
\end{eqnarray}
The role of the mediator anharmonicity becomes transparent by
introducing the interaction-induced anharmonicity of the battery
ladder, $\Delta_B \equiv \Omega_{B,2}-\Omega_{B,1}.$ The simultaneous resonance conditions then imply
\begin{eqnarray}
		\alpha_M=-\Delta_B.
	\label{eq:complementary_anharmonicity}
\end{eqnarray}
Thus the mediator anharmonicity compensates the
interaction-induced nonequidistance of the battery spectrum.
Whenever $\Omega_{B,1}\neq\Omega_{B,2}$, a harmonic mediator
($\alpha_M=0$) cannot simultaneously match both battery
transitions.  The anharmonic mediator therefore provides the
spectral structure required to establish the two cross-matched resonant
$M$--$B$ transfer channels.\\

\subsubsection{Effective mediator--battery Hamiltonian}

During the $M$--$B$ charging stage, the relevant two-excitation
subspace is
\begin{eqnarray}
	\mathcal H_{MB}^{(2)}
	=
	{\rm span}
	\left\{
	\ket{2_M,00_B},
	\ket{1_M,B_+},
	\ket{0_M,11_B}
	\right\}.
\end{eqnarray}
%
In a rotating frame, the effective Hamiltonian can be
written as
\begin{eqnarray}
	H_{MB,{\rm eff}}^{(2)}
	=
	\left(
	\begin{array}{ccc}
		0
		&
		2g(t)
		&
		0
		\\
		2g(t)
		&
		-\delta_{MB}^{(2,1)}
		&
		\sqrt{2}g(t)
		\\
		0
		&
		\sqrt{2}g(t)
		&
		-\delta_{MB}^{(2,1)}
		-\delta_{MB}^{(1,2)}
	\end{array}
	\right).
	\label{eq:HMB_eff_detuned}
\end{eqnarray}
At the double-resonance condition in
Eq.~\eqref{eq:double_MB_resonance},
\begin{eqnarray}
	H_{MB,{\rm eff}}^{(2)}
	=
	g(t)
	\left(
	\begin{array}{ccc}
		0 & 2 & 0\\
		2 & 0 & \sqrt{2}\\
		0 & \sqrt{2} & 0
	\end{array}
	\right).
	\label{eq:HMB_eff_resonant}
\end{eqnarray}
Since the time dependence appears only through the common envelope
$g(t)$, Hamiltonians at different times commute.  Introducing
\begin{eqnarray}
	\Theta_g(t)
	=
	\int_0^t g(t')\,dt',
	\label{eq:pulsearea}
\end{eqnarray}
the target-state population for an initial
$\ket{2_M,00_B}$ state follows exactly as
\begin{eqnarray}
	P_{11_B}(t) 
	= \norm{ A_{\ket{2_M,00_B} \rightarrow \ket{0_M,11_B}}(t) }^2 = 
	\frac{8}{9}
	\sin^4
	\left[
	\frac{\sqrt{6}}{2}
	\Theta_g(t)
	\right].
	\label{eq:P11_exact}
\end{eqnarray}
Leading to maximum transfer probability $P_{11}^{\rm max} = \frac{8}{9}.$ The role of $\alpha_M$ is to satisfy the two
simultaneous resonance conditions in
Eq.~\eqref{eq:double_MB_resonance}; once resonance is established,
the remaining $8/9$ ceiling originates solely from the unequal
matrix elements $2g$ and $\sqrt{2}g$ of the three-state chain.

\subsection{Conditions for perfect two-excitation transfer}

The efficiency of resource transfer through the $S$--$M$--$B$
architecture is jointly controlled by excitation-sector
accessibility, spectral matching, transition matrix elements, and
interaction pulse areas.  
For an isolated resonant two-state
exchange, maximum population transfer occurs whenever
\begin{eqnarray}
	\Theta_{\rm eff}
	=
	\frac{(2n+1)\pi}{2},
	\qquad
	n=0,1,\ldots .
	\label{eq:pi_swap_general}
\end{eqnarray}
%




Thus for a general resonant three-state chain
\begin{eqnarray}
	H_{\rm eff}
	=
	\left(
	\begin{array}{ccc}
		0&G_1&0\\
		G_1&0&G_2\\
		0&G_2&0
	\end{array}
	\right),
\end{eqnarray}
the maximum end-to-end transfer probability is
\begin{eqnarray}
	P_{\rm max}
	=
	\frac{4|G_1|^2|G_2|^2}
	{\left(|G_1|^2+|G_2|^2\right)^2}.
	\label{eq:Pmax_general_chain}
\end{eqnarray}
The two effective couplings $G_1$ and $G_2$ characterize the
strengths of the consecutive links of the transfer chain.  An
imbalance between them limits end-to-end population transport; we
refer to the removal of this coupling mismatch as `impedance
matching'. Unit transfer therefore requires the impedance-matching condition 
\begin{eqnarray}
	|G_1|=|G_2|.
	\label{eq:immatching}
\end{eqnarray}
 The original interaction in Eq.~\eqref{eq:HMB_eff_resonant}
instead gives
$G_1=2g$ and $G_2=\sqrt{2}g$, producing the 8/9 ceiling. The mismatch can be compensated by
a transition-resolved mediator coupling,
\begin{eqnarray}
	H_{MB}^{\rm eng}(t)
	=
	g(t)
	\left[
	\eta_{21}\sqrt{2}
	\ket{1_M}\bra{2_M}
	+
	\eta_{10}
	\ket{0_M}\bra{1_M}
	\right]
	\left(
	\sigma_+^{B_1}+\sigma_+^{B_2}
	\right)
	+
	{\rm H.c.},
	\label{eq:HMB_engineered}
\end{eqnarray}
for which $G_1 = 2\eta_{21}g$, $G_2 = \sqrt{2}\eta_{10}g$.  Eq.~\eqref{eq:immatching} is then satisfied by $\eta_{10} = \sqrt{2}\eta_{21} $. Under simultaneous spectral matching condition
Eq.~\eqref{eq:double_MB_resonance} and matrix-element or impedance matching condition Eq.~\eqref{eq:immatching}, the effective chain becomes
mirror symmetric and supports $\ket{2_M,00_B} \longrightarrow \ket{0_M,11_B}$ with unit probability at the appropriate pulse area defined by Eq.~\eqref{eq:pulsearea}.\\

Consequently, deterministic two-excitation charging requires two
conceptually distinct ingredients.  The first is spectral matching, which makes both excitation-transfer channels accessible. The second is impedance matching for simultaneous transfer to obtain $P_{11}=1$.

\section{Charging dynamics and resource transfer}
\label{sec:numerics}  

For a specified mediator and battery preparation, the transfer starts
from
\begin{eqnarray}
	\rho_{SMB}(0)
	=
	\rho_S^{(e)}
	\otimes
	\rho_M(0)
	\otimes
	\rho_B(0),
\end{eqnarray}
and ends at
\begin{eqnarray}
	\rho_{SMB}^{({\rm tr})}
	=
	U_{\rm tr}\rho_{SMB}(0)U_{\rm tr}^{\dagger},
	\qquad
	U_{\rm tr}
	=
	{\cal T}
	\exp\!\left[
	-i\int_0^{\tau_{\rm tr}}
	dt\,H_{\rm SMB}(t)
	\right].
	\label{eq:rhoSMB_final}
\end{eqnarray}
The corresponding reduced states are
\begin{eqnarray}
	\rho_S^{({\rm tr})}
	&=&
	{\rm Tr}_{MB}
	\left[
	\rho_{SMB}^{({\rm tr})}
	\right],
	\qquad
	\rho_M^{({\rm tr})}
	=
	{\rm Tr}_{SB}
	\left[
	\rho_{SMB}^{({\rm tr})}
	\right], 
	\qquad
	\rho_B^{({\rm tr})}
	=
	{\rm Tr}_{SM}
	\left[
	\rho_{SMB}^{({\rm tr})}
	\right].
	\label{eq:reduced_transfer_states}
\end{eqnarray}

For the numerical results demonstrating the charging dynamics and
resource transfer, we use
\begin{eqnarray}
\beta=5,\qquad
\Delta_S=1.3,\qquad
\xi_h=4,\qquad
\xi_c=1,\qquad
\tau_e=1,
\label{eq:num_benchmark1}
\end{eqnarray}
which yields
$\omega_S^{(c)}\simeq1.640$ and the expansion-generated resource
measures
$P_e\simeq0.0524$,
$C_S^{(e)}\simeq0.445$, and
${\cal W}_e\simeq0.0859$.
Unless stated otherwise, we further take
\begin{eqnarray}
J_0=0.20,\qquad
\alpha_M=-0.10,\qquad
\lambda_B=0.01,\qquad
\chi_B=0.03,
\label{eq:num_benchmark2}
\end{eqnarray}
which satisfy
$\alpha_M=2(\lambda_B-2\chi_B)$
[Eq.~\eqref{eq:double_res_parameters}] and thereby permit simultaneous
cross matching of the mediator and battery transitions in the two-excitation sector.\\

For an initially empty battery,  
convenient bases for the accessible
one- and two-excitation sectors are 
\begin{eqnarray}
{\cal B}_{1}
&=&
\Big\{
\ket{\varepsilon_+^{(c)},0_M,00_B},
\ket{\varepsilon_-^{(c)},1_M,00_B},
\ket{\varepsilon_-^{(c)},0_M,B_+}
\Big\},
\nonumber\\
{\cal B}_{2}
&=&
\Big\{
\ket{\varepsilon_+^{(c)},1_M,00_B},
\ket{\varepsilon_-^{(c)},2_M,00_B},
\ket{\varepsilon_-^{(c)},1_M,B_+},
\ket{\varepsilon_-^{(c)},0_M,11_B}
\Big\}.
\label{eq:smbbasis}
\end{eqnarray}
For $\ket{\Phi}\in{\cal B}_n$, we define the conditional population
$\widetilde P_{\Phi}(t)$ by
\begin{eqnarray}
\bra{\Phi}
\rho_{SMB}(t)
\ket{\Phi}
=
p_+^{(e)}
\widetilde P_{\Phi}(t),
\qquad
p_+^{(e)}
=
\bra{\varepsilon_+^{(c)}}
\rho_S^{(e)}
\ket{\varepsilon_+^{(c)}}.
\label{eq:conditional_population}
\end{eqnarray}
Because the mediator and battery are initialized with definite
excitation number, $p_+^{(e)}$ gives the weight of the corresponding
active excitation sector and is conserved by ${\cal N}$.
Accordingly, $\widetilde P_{\Phi}(t)$ directly tracks the instantaneous redistribution
of population within that sector. \\

\begin{figure*}[t]
\centering
\begin{subfigure}[t]{0.48\textwidth}
\centering
\includegraphics[width=\linewidth]{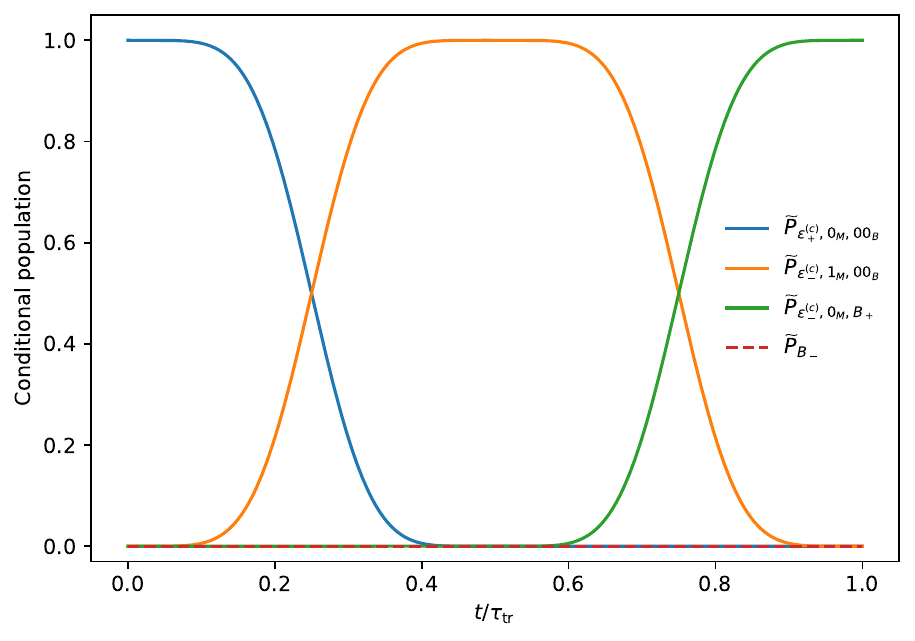}
\caption{}
\label{fig:pathwaysa}
\end{subfigure}
\hfill
\begin{subfigure}[t]{0.48\textwidth}
\centering
\includegraphics[width=\linewidth]{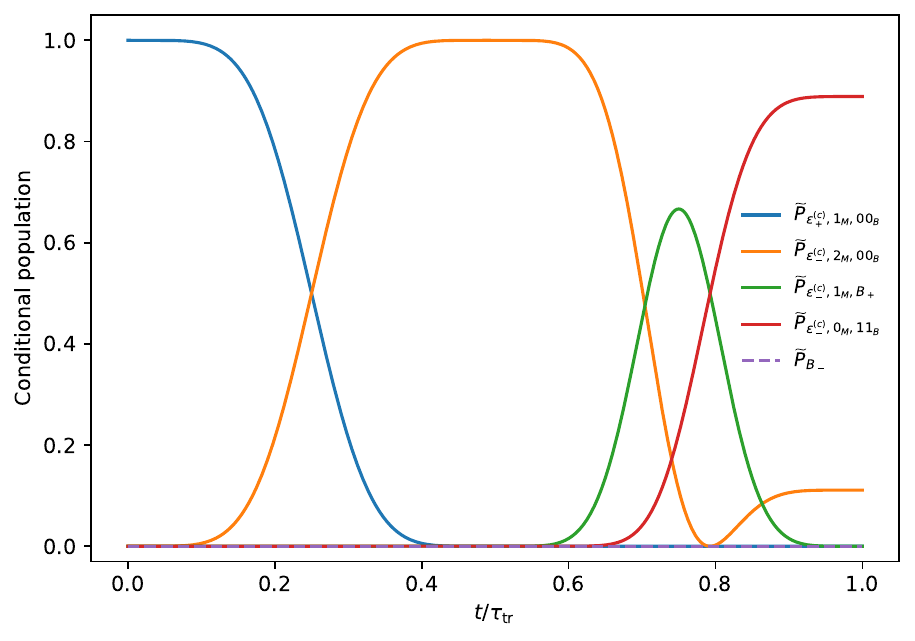}
\caption{}
\label{fig:pathwaysb}
\end{subfigure}
\caption{\textit{Conditional excitation-transfer dynamics.}
(a) Single-excitation transfer for
$\rho_M(0)=\ket{0_M}\bra{0_M}$ and
$\rho_B(0)=\ket{00_B}\bra{00_B}$ under
$\omega_S^{(c)}=\Omega_{M,1}=\Omega_{B,1}$.
We use
$\omega_M=1.64$, $\omega_B=1.69$,
$J_0=0.20$, $g_0=J_0/\sqrt{2}$, and
$\tau_{\rm tr}=2\pi/J_0\simeq31.42$.
(b) Two-excitation transfer for
$\rho_M(0)=\ket{1_M}\bra{1_M}$ and
$\rho_B(0)=\ket{00_B}\bra{00_B}$ under
$\omega_S^{(c)}=\Omega_{M,2}=\Omega_{B,1}$ and
$\Omega_{M,1}=\Omega_{B,2}$.
Here
$\omega_M=1.74$, $\omega_B=1.69$,
$g_0=(2/\sqrt{3})J_0$, and
$\tau_{\rm tr}=\sqrt{2}\pi/J_0\simeq22.21$.
Populations are normalized by $p_+^{(e)}$; the dark-state occupation
remains negligible.}
\label{fig:pathways}
\end{figure*}
Fig.~\ref{fig:pathways} resolves the two charging pathways
for parameter values given in  the figure caption. For an unloaded mediator
[Fig.~\ref{fig:pathwaysa}], the resonance condition
$\omega_S^{(c)}=\Omega_{M,1}=\Omega_{B,1}$ and matched pulse areas drive the transition
\begin{eqnarray}
\ket{\varepsilon_+^{(c)},0_M,00_B}
\rightarrow
\ket{\varepsilon_-^{(c)},1_M,00_B}
\rightarrow
\ket{\varepsilon_-^{(c)},0_M,B_+},
\end{eqnarray}
yielding an almost complete, self-resetting transfer with the mediator
returning to $\ket{0_M}$.
For an initially loaded mediator [Fig.~\ref{fig:pathwaysb}], the
cross-resonance conditions
$\omega_S^{(c)}=\Omega_{M,2}=\Omega_{B,1}$ and
$\Omega_{M,1}=\Omega_{B,2}$ activate the cascade
\begin{eqnarray}
\ket{\varepsilon_+^{(c)},1_M,00_B}
\rightarrow
\ket{\varepsilon_-^{(c)},2_M,00_B}
\rightarrow
\ket{\varepsilon_-^{(c)},1_M,B_+}
\rightarrow
\ket{\varepsilon_-^{(c)},0_M,11_B}.
\end{eqnarray}
The target-state population approaches $8/9$, with the residual
$\simeq1/9$ remaining mainly in
$\ket{\varepsilon_-^{(c)},2_M,00_B}$. This incomplete transfer persists
at exact resonance and  follows from the
unequal $2g(t)$ and $\sqrt{2}g(t)$ couplings of the effective
three-state chain~\cite{Christandl2004}.  \\

\begin{figure*}[t]
\centering
\begin{subfigure}[t]{0.48\textwidth}
\centering
\includegraphics[width=\linewidth]{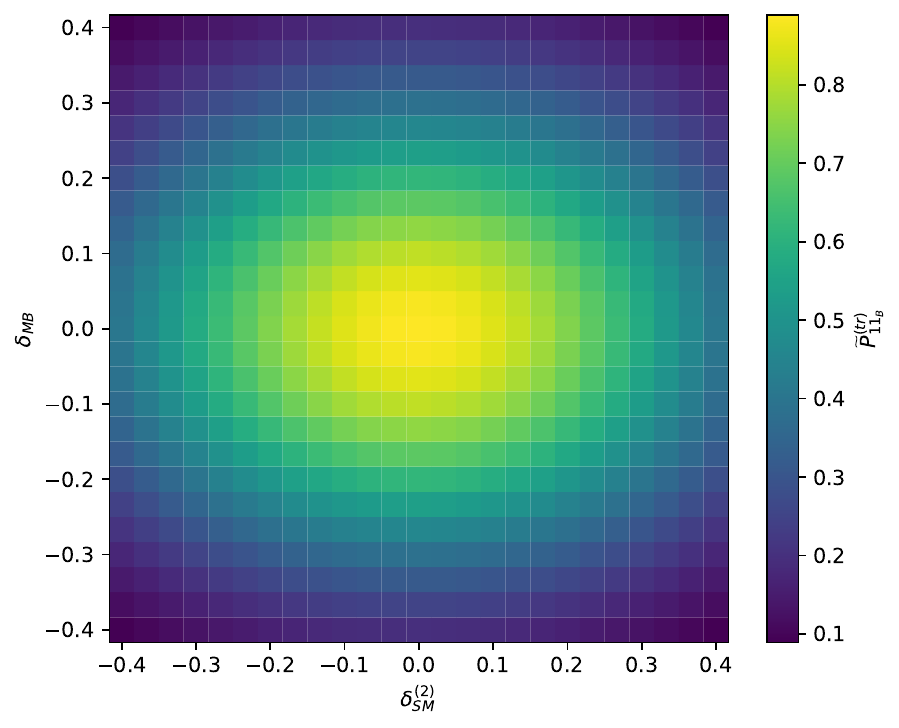}
\caption{}
\label{fig:res-map}
\end{subfigure}
\hfill
\begin{subfigure}[t]{0.48\textwidth}
\centering
\includegraphics[width=\linewidth]{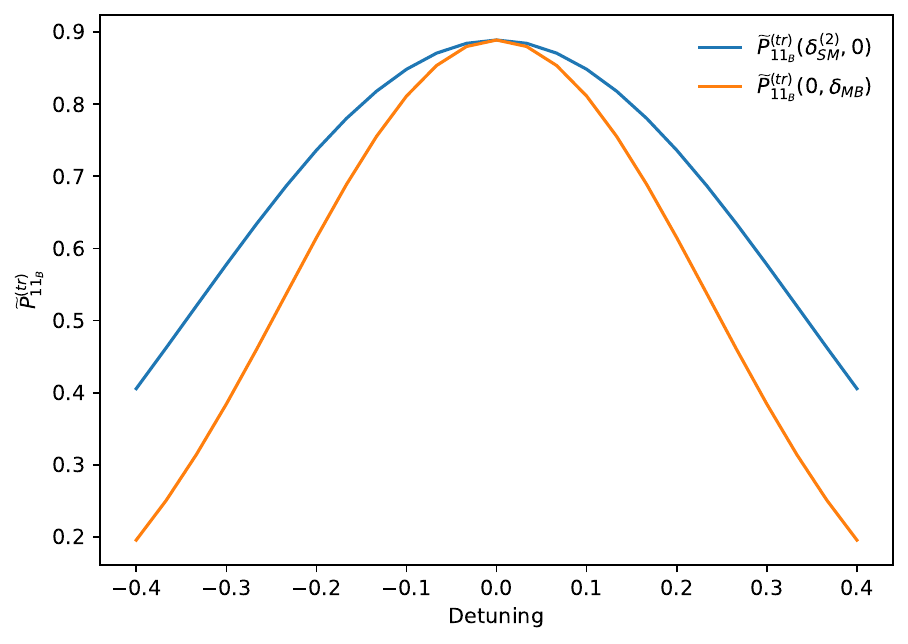}
\caption{}
\label{fig:res-cuts}
\end{subfigure}
\caption{\textit{Resonance sensitivity of the two-excitation channel.}
(a) Final conditional population of $\ket{11_B}$ in the
$(\delta_{SM}^{(2)},\delta_{MB})$ plane, where
$\delta_{MB}\equiv\delta_{MB}^{(2,1)}
=\delta_{MB}^{(1,2)}$.
The detunings are varied over $[-0.4,0.4]$ for
$\alpha_M=-0.10$, $\lambda_B=0.01$, $\chi_B=0.03$,
$J_0=0.20$, $g_0=(2/\sqrt{3})J_0$, and
$\tau_{\rm tr}=\sqrt{2}\pi/J_0\simeq22.21$.
(b) Line cuts obtained by fixing one detuning at resonance while
scanning the other.}
\label{fig:resonance}
\end{figure*}

To quantify the spectral sensitivity of the loaded two-excitation
cascade, we vary the detunings
$\delta_{SM}^{(2)}$ and
$\delta_{MB}\equiv
\delta_{MB}^{(2,1)}=\delta_{MB}^{(1,2)}$.
The final conditional population of the doubly excited battery state,
\begin{eqnarray}
\widetilde P_{11_B}^{(\rm tr)}
=
\frac{
\bra{11_B}\rho_B^{(\rm tr)}\ket{11_B}
}{
p_+^{(e)}
},
\label{eq:p11bfinal}
\end{eqnarray}
directly measures transfer into
$\ket{\varepsilon_-^{(c)},0_M,11_B}$ within the conserved
two-excitation sector. \\

\begin{figure}[t]
	\centering
	\begin{subfigure}[t]{0.48\textwidth}
		\centering
		\includegraphics[width=\linewidth]{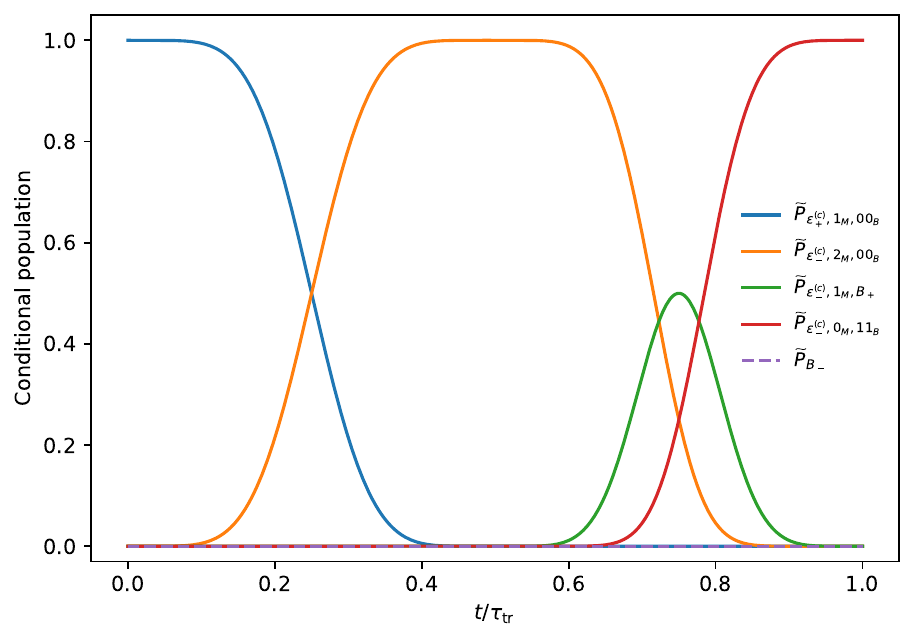}
		\caption{}
		\label{fig:pathwayeta}
	\end{subfigure}
	\hfill
	\begin{subfigure}[t]{0.48\textwidth}
		\centering
		\includegraphics[width=\linewidth]{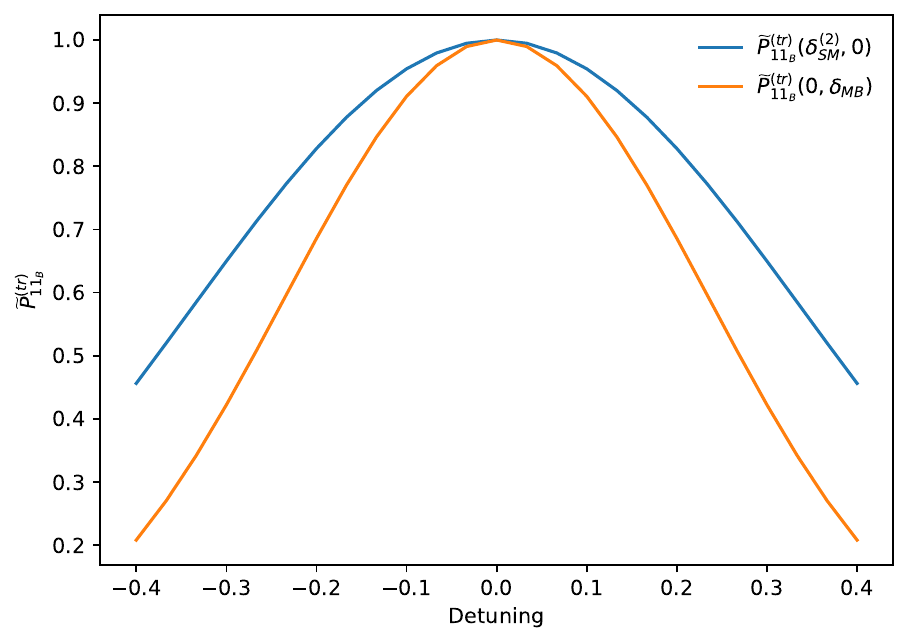}
		\caption{}
		\label{fig:res-cuts-eta}
	\end{subfigure}
		\caption{(a)Two-excitation transfer for $\rho_M(0)=\ket{1_M}\bra{1_M}$ and $\rho_B(0)=\ket{00_B}\bra{00_B}$ at $\omega_S^{(c)}=\Omega_{M,2}=\Omega_{B,1}$ and $\Omega_{M,1}=\Omega_{B,2}$.
		Parameters are $\eta_{21}=1/\sqrt{2}$, $\eta_{10}=1$, $\omega_M\simeq1.74$, $\omega_B\simeq1.69$, $J_0=0.20$, $g_0=\sqrt{2}J_0$, and $\tau_{\rm tr}=\sqrt{2}\pi/J_0\simeq22.21$.
		Populations are normalized by $p_+^{(e)}$, yielding $\widetilde P_{11_B}(\tau_{\rm tr})=1$; the dark-state occupation remains negligible. (b) Line cuts obtained by fixing one detuning at resonance while
		scanning the other.
	}
	
\end{figure}
Fig.~\ref{fig:pathwayeta} demonstrates complete conditional two-excitation transfer under double resonance. The engineered weights $\eta_{21}=1/\sqrt{2}$ and $\eta_{10}=1$ equalize the cascade couplings, $G_1(t)=G_2(t)=\sqrt{2}g(t)$, removing the $8/9$ transfer bound of the unengineered interaction. With $g_0=\sqrt{2}J_0$ and $\tau_{\rm tr}=\sqrt{2}\pi/J_0$, the common mediator--battery pulse yields $\widetilde P_{11_B}(\tau_{\rm tr})=1$, consistent with the numerical evolution. The unnormalized final population is $P_{11_B}=p_+^{(e)}$.\\

As shown in Fig.~\ref{fig:resonance} and Fig.~\ref{fig:res-cuts-eta}, the resonance profiles at
$(\delta_{SM}^{(2)},\delta_{MB})=(0,0)$ with
$\widetilde P_{11_B}^{(\rm tr)}\simeq8/9$ and 1 respectively, demonstrating that efficient double charging requires simultaneous matching of the $S$--$M$ and
$M$--$B$ links.

The population decreases smoothly upon detuning.
The corresponding line cuts in Fig.~\ref{fig:res-cuts}, obtained by fixing one detuning at resonance and scanning the other, yield the widths 
\begin{eqnarray}
{\rm FWHM}_{SM}\simeq0.75,
\qquad
{\rm FWHM}_{MB}\simeq0.55 ,
\end{eqnarray}
showing a narrower, and hence more detuning-sensitive, $M$--$B$ resonance for the benchmark specified in the figure caption.\\

To quantify the transfer of energetic resources, we evaluate the changes in
internal energy and ergotropy 
\begin{eqnarray}
\Delta E_X
&=&
{\rm Tr}
\left[
H_X
\left(
\rho_X^{({\rm tr})}-\rho_X(0)
\right)
\right],
\qquad
\Delta{\cal W}_X
=
{\cal W}
\left[
\rho_X^{({\rm tr})},H_X
\right]
-
{\cal W}
\Big[
\rho_X(0),H_X
\Big] 
\label{eq:resource_transfer}
\end{eqnarray}
of each subsystem $X=S,M,B$ over the transfer
stroke of duration $\tau_{\rm tr}$.
Energy and ergotropy are tracked separately, since unitary redistribution of
local coherence and correlations can render part of the deposited energy
passive and hence nonextractable~\cite{Allahverdyan2004,
Andolina2019Ergotropy,Touil2021}.\\

\begin{figure}[t]
\centering
\begin{subfigure}[t]{0.49\textwidth}
\centering
\includegraphics[width=\linewidth]{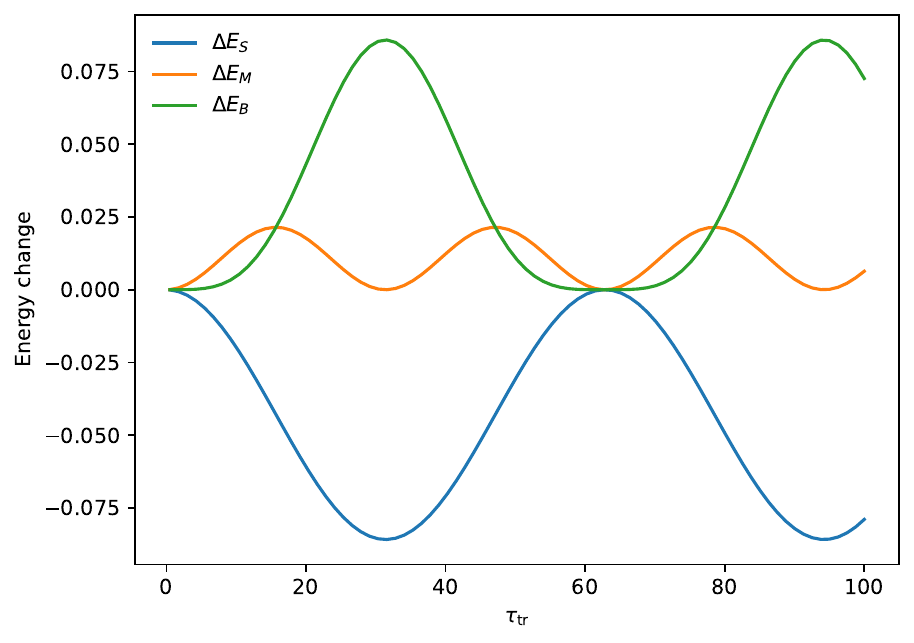}
\caption{Ground mediator: energy transfer.}
\label{fig:g-energy}
\end{subfigure}
\hfill
\begin{subfigure}[t]{0.49\textwidth}
\centering
\includegraphics[width=\linewidth]{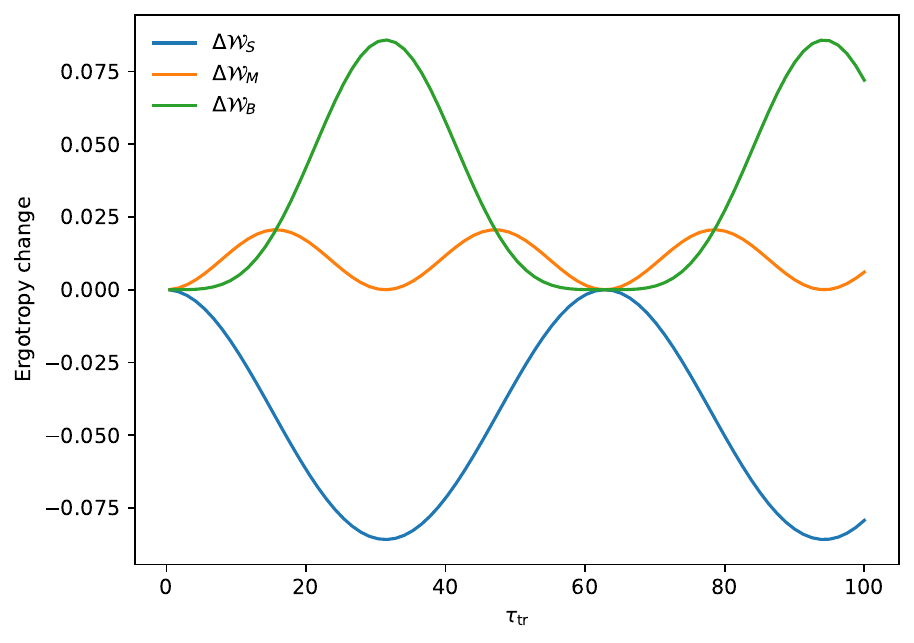}
\caption{Ground mediator: ergotropy transfer.}
\label{fig:g-erg}
\end{subfigure}

\vspace{0.5em}

\begin{subfigure}[t]{0.49\textwidth}
\centering
\includegraphics[width=\linewidth]{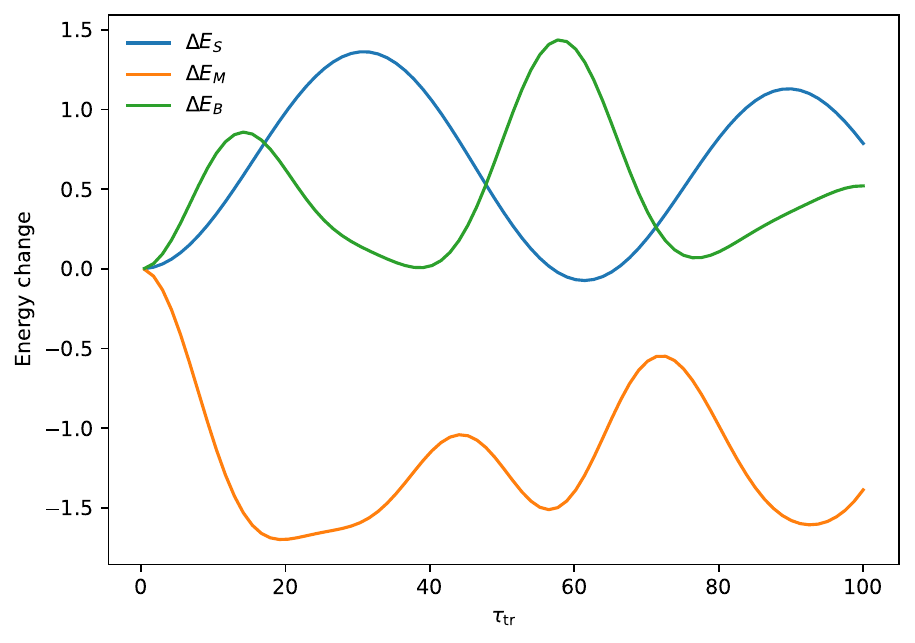}
\caption{Loaded mediator: energy transfer.}
\label{fig:l-energy}
\end{subfigure}
\hfill
\begin{subfigure}[t]{0.49\textwidth}
\centering
\includegraphics[width=\linewidth]{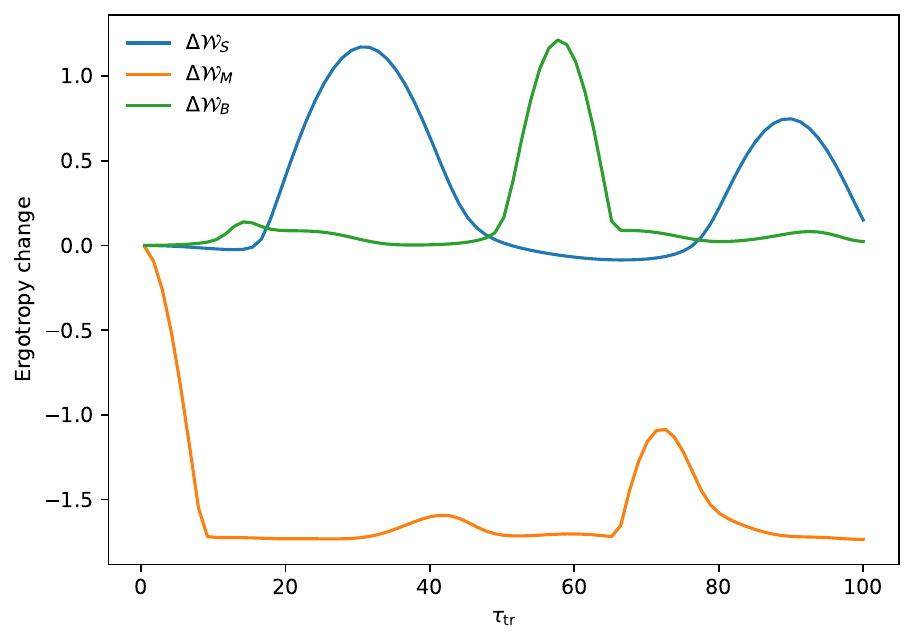}
\caption{Loaded mediator: ergotropy transfer.}
\label{fig:l-erg}
\end{subfigure}
\caption{\textit{Resource redistribution during the transfer stroke.}
Panels (a) and (b) show local energy and ergotropy changes for the
resonant unloaded-mediator protocol with
$\omega_M=1.64$, $\omega_B=1.69$,
$J_0=0.20$, and $g_0=J_0/\sqrt{2}$.
Panels (c) and (d) show the corresponding loaded-mediator protocol with
$\omega_M=1.74$, $\omega_B=1.69$,
$J_0=0.20$, and $g_0=(2/\sqrt{3})J_0$.
The input state is fixed by Eq.~\eqref{eq:num_benchmark1}, and
$0.5\leq\tau_{\rm tr}\leq100$.
Unlike Fig.~\ref{fig:pathways}, these quantities are evaluated from the
complete mixed state and are not normalized by $p_+^{(e)}$.}
\label{fig:resource-scans}
\end{figure}

For the unloaded mediator,
Figs.~\ref{fig:g-energy} and \ref{fig:g-erg} show nearly coincident
energy- and ergotropy-transfer profiles. 
At the first complete $S \to M \to B$ transfer,
$\tau_{\rm swap}^{(g,1)}
=2\pi/J_0\simeq31.42$,
the numerical maximum occurs near
$\tau_{\rm tr}\simeq31.7$ with
\begin{eqnarray}
\Delta E_B
\simeq
\Delta{\cal W}_B
\simeq0.0858,
\qquad
\Delta E_M,
\Delta{\cal W}_M
\sim2\times10^{-5},
\label{eq:ground_resource_swap}
\end{eqnarray}
showing nearly complete deposition of the extracted resource in the
battery as ergotropy, with negligible residual storage in the mediator. The battery returns close to its initial resource content near
$\tau_{\rm tr}\simeq4\pi/J_0$, and the second end-to-end charging
maximum occurs at
$\tau_{\rm swap}^{(g,2)}
=6\pi/J_0\simeq94.25$, reflecting coherent recurrence of the
single-excitation transfer.\\

The transfer dynamics for an initially loaded mediator,
shown in Figs.~\ref{fig:l-energy} and~\ref{fig:l-erg}, is qualitatively
distinct because $M$ both mediates the exchange and supplies its initially
stored excitation energy to the charging process. 
The first pulse-area-matched transfer in the resonant two-excitation
cascade occurs at
$\tau_{\rm swap}^{(l,1)}=\sqrt{2}\pi/J_0\simeq22.21$. As discussed above,
the unequal $2g(t)$ and $\sqrt{2}g(t)$ couplings limit the conditional
population of $\ket{11_B}$ to $8/9$. 
Near this time $\tau_{\rm tr} \simeq 22.1$, 
the transfer dynamics computation gives 
\begin{eqnarray}
\Delta E_S\simeq1.10,\qquad
\Delta E_M\simeq-1.68,\qquad
\Delta E_B\simeq0.49,\qquad
\Delta{\cal W}_B\simeq0.086.
\end{eqnarray}
The large difference between $\Delta E_B$ and
$\Delta{\cal W}_B$ shows that a substantial fraction of the deposited
battery energy is passive, while $\Delta E_M<0$ identifies the mediator
as an energetic contributor in this protocol. 
Because the full initial state contains both ground- and excited-state
components of $S$, the complete mixed-state dynamics contains more than
the conditional active branch. Its resource extrema therefore need not
coincide with the nominal sector-resolved swap times
~\cite{Bialczak2011,Sete2015,Blais2021CircuitQED}. This is evident from
the pronounced battery maximum near
\begin{eqnarray}
\tau_{\rm tr}
\simeq57.8,
\qquad
\Delta E_B
\simeq1.44,
\qquad
\Delta{\cal W}_B
\simeq1.21,
\label{eq:loaded_second_peak}
\end{eqnarray}
whereas the next nominal two-excitation recurrence occurs at
$\tau_{\rm swap}^{(l,2)}
=3\sqrt{2}\pi/J_0\simeq66.64$.
The transfer duration is therefore an independent control parameter
for the full multilevel charging dynamics.

\section{Compression and cycle closure}
\label{sec:transferandclosure}
After the transfer stroke, 
$S$ is decoupled from $M$ and $B$ and
compressed from $\xi_c$ to $\xi_h$ over a duration $\tau_c$ using
\begin{eqnarray}
\xi(t)
=
\xi_c
+
(\xi_h-\xi_c)
f\!\left(\frac{t}{\tau_c}\right),
\qquad
0\leq t\leq\tau_c.
\end{eqnarray}
Its state becomes
\begin{eqnarray}
\rho_S^{({\rm comp})}
=
U_c
\rho_S^{({\rm tr})}
U_c^\dagger,
\qquad
U_c
=
{\cal T}
\exp\!\left[
-i\int_0^{\tau_c}
dt\,H_S(t)
\right],
\label{eq:rho_comp}
\end{eqnarray}
and the work performed on $S$ is
\begin{eqnarray}
W_c
&=&
{\rm Tr}
\left[
\rho_S^{({\rm comp})}
H_S^{(h)}
\right]
-
{\rm Tr}
\left[
\rho_S^{({\rm tr})}
H_S^{(c)}
\right].
\label{eq:W_compression}
\end{eqnarray}
Rethermalization at fixed $H_S^{(h)}$ restores
$\rho_S^{(h),{\rm G}}$ and closes the working-system cycle.\
Taking heat to be positive when absorbed by $S$, the heat
exchanged during the thermal reset is  
\begin{eqnarray}
Q
&=&
{\rm Tr}
\left[
H_S^{(h)}
\left(
\rho_S^{(h),{\rm G}}
-
\rho_S^{({\rm comp})}
\right)
\right].
\label{eq:Q_cycle}
\end{eqnarray}
Because the transfer couplings are externally modulated, their switching
contributes an additional work term. Within the standard driven-Hamiltonian
decomposition~\cite{Alicki1979,Campisi2011,Carrega2022}, this switching
work is
\begin{eqnarray}
W_{\rm switch}
&=&
\int_0^{\tau_{\rm tr}} dt\,
{\rm Tr}\!\left[
\rho_{SMB}(t)\,
\dot H_{\rm int}(t)
\right],
\qquad
H_{\rm int}(t)
=
H_{SM}(t)+H_{MB}(t).
\label{eq:Wswitch}
\end{eqnarray}
For the nonoverlapping protocol in Eq.~\eqref{eq:Jg},
$W_{\rm switch}
=
W_{\rm switch}^{SM}
+
W_{\rm switch}^{MB}$.
Since $S$ returns to its initial Gibbs state after the reset, the
cycle-level balance is
\begin{eqnarray}
Q
+
W_e
+
W_c
+
W_{\rm switch}
=
\Delta E_B+\Delta E_M .
\label{eq:cycle_closure}
\end{eqnarray}
In the unloaded-mediator protocol
$\Delta E_M\simeq0$, so the mediator functions primarily as a coherent
transport channel rather than as a consumable energetic resource.\\

\begin{figure*}[t]
\centering
\begin{subfigure}[t]{0.48\textwidth}
\centering
\includegraphics[width=\linewidth]{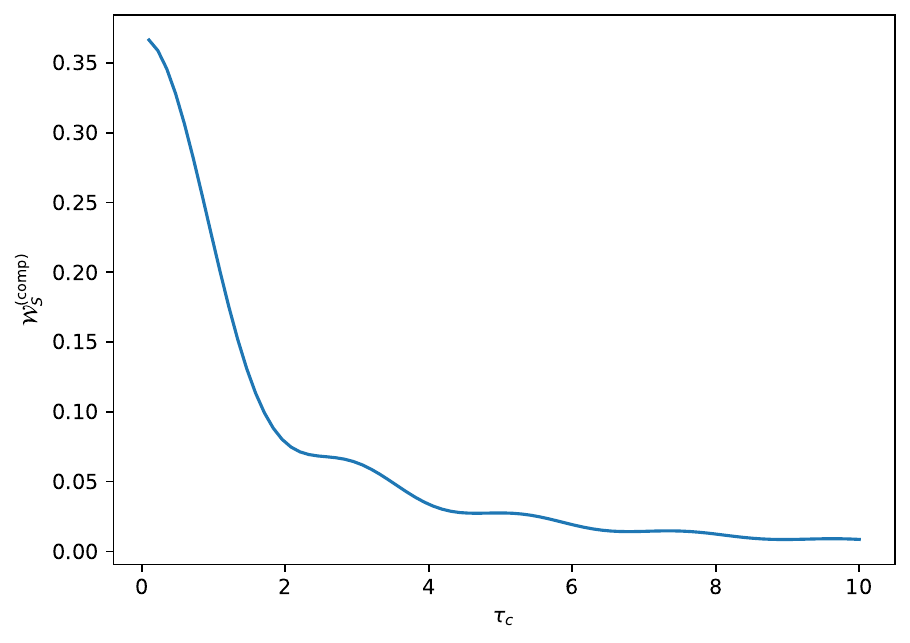}
\caption{Residual working-system ergotropy.}
\label{fig:comp-erg}
\end{subfigure}
\hfill
\begin{subfigure}[t]{0.48\textwidth}
\centering
\includegraphics[width=\linewidth]{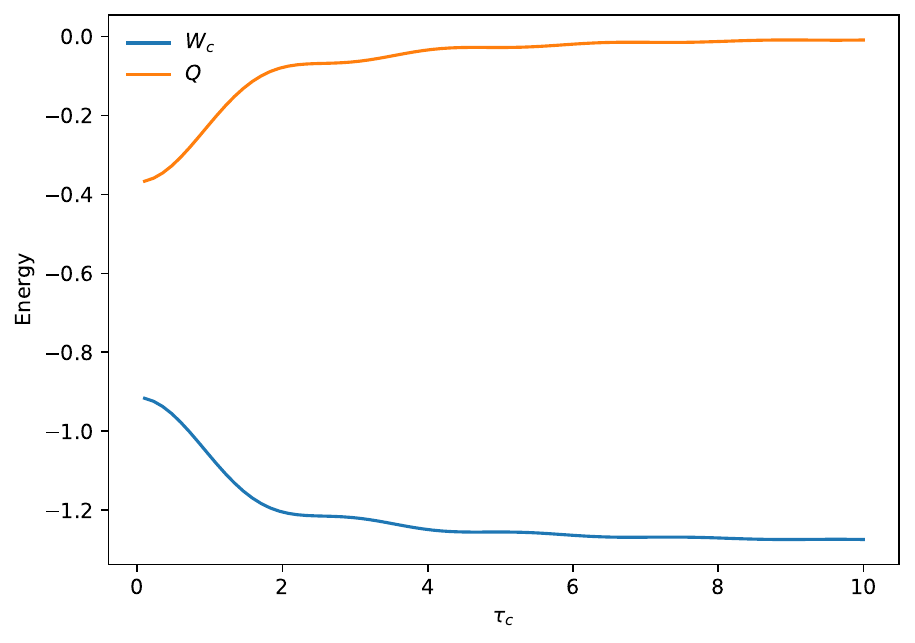}
\caption{Compression work and reset heat.}
\label{fig:comp-work}
\end{subfigure}
\caption{\textit{Finite-time compression and cycle closure.}
The compression starts from the first near-optimal unloaded-mediator
transfer point,
$\omega_M=1.64$, $\omega_B=1.69$,
$J_0=0.20$, $g_0=J_0/\sqrt{2}$, and
$\tau_{\rm tr}=2\pi/J_0$.
The control is returned from $\xi_c=1$ to $\xi_h=4$ over
$0.1\leq\tau_c\leq10$.
(a) Residual ergotropy
${\cal W}_S^{({\rm comp})}$ after compression.
(b) Compression work $W_c$ and heat $Q$ exchanged during the
subsequent reset. Work and heat are positive when energy is supplied
to $S$.}
\label{fig:compression}
\end{figure*}

Fig.~\ref{fig:compression} shows the return stroke from the first
near-optimal unloaded-mediator charging point,
$\tau_{\rm tr}=2\pi/J_0\simeq31.42$, where the mediator is nearly reset
and the battery stores most of the expansion-generated ergotropy. The
resulting state of $S$ is then compressed over a duration $\tau_c$.
The compression time controls the residual nonequilibrium of $S$ and,
through $\rho_S^{({\rm comp})}$, the heat exchanged during the
subsequent thermal reset.  
As shown in Fig.~\ref{fig:comp-erg}, the residual ergotropy
${\cal W}_S^{({\rm comp})}$ decreases from $\simeq0.366$ at
$\tau_c=0.1$ to $\simeq8.6\times10^{-3}$ at $\tau_c=10$, indicating that
slower compression leaves $S$ progressively closer to a passive state. 
 Correspondingly, $Q<0$ throughout
[Fig.~\ref{fig:comp-work}], indicating heat release to the reservoir,
with $|Q|$ decreasing over the same range. The compression work $W_c$ is
also negative, implying energy recovery by the external drive, and
approaches its slow-driving limit as $\tau_c$ increases.

Finally, defining the numerical cycle residual as
\begin{eqnarray}
\epsilon_{\rm cyc}
=
Q + W_e + W_c + W_{\rm switch}
- \Delta E_B - \Delta E_M,
\label{eq:cycle_residual}
\end{eqnarray}
we obtain $\displaystyle
\max_{\tau_c}
\left|
\epsilon_{\rm cyc}
\right|
\simeq
7.3\times10^{-16}$ across the compression scan,
confirming closure of the expansion, transfer, compression, and
thermal-reset energy balance to numerical precision.

\section{Conclusion}
\label{sec:conclusion}

We have investigated a finite-time charging architecture in which a
driven two-level working system transfers nonequilibrium resources to
an interacting two-qubit battery through a weakly anharmonic
three-level mediator. Finite-time expansion generates coherence and
ergotropy in the working system, while excitation-number conservation
organizes the subsequent transfer into spectrally distinct sectors. The mediator plays two conceptually distinct roles in this
architecture. Beyond providing an indirect transfer channel, the mediator introduces a nonuniform transition spectrum that can be matched to the interaction-modified excitation spectrum of the battery.
\\

The single-excitation sector supports nearly complete $S\rightarrow M\rightarrow B$ transfer through the lowest mediator transition. In the two-excitation sector, both mediator transitions participate, and simultaneous resonance with the unequal battery gaps requires $\alpha_M = \Omega_{B,1}-\Omega_{B,2}$. At exact double resonance, the $M$--$B$ dynamics reduces to an effective three-state chain. For the natural couplings $G_1=2g$ and $G_2=\sqrt{2}g$, the unequal matrix elements impose the exact bound $P_{\rm max}=8/9$. Unit transfer is restored when the impedance-matching condition $|G_1|=|G_2|$ is satisfied, in agreement with the numerical dynamics. The detuning dependence sets the corresponding spectral tolerance, while the full mixed-state dynamics shows that battery-energy
and ergotropy maxima can depart from the sector-resolved swap times. In the loaded-mediator
protocol, part of the energy deposited in the battery originates from
the mediator, and a substantial fraction of this contribution may be
stored as passive energy rather than ergotropy. \\

These results separate two requirements for efficient multiexcitation charging: spectral matching determines whether the relevant transfer channels are simultaneously accessible, while impedance matching determines whether those channels support complete population transfer. The mediator therefore acts as a spectral interface whose level structure and transition matrix elements jointly control charging performance, providing a natural route toward collective and multilevel battery architectures.\\

Finally, including finite-time compression, thermal reset, and the work
associated with modulation of the transfer couplings provides a complete
cycle-level energy accounting, with the numerical energy-balance
residual remaining at the level of machine precision. The resulting
framework therefore connects finite-time resource generation, coherent
multilevel transfer, mediator recycling, and thermodynamic cycle closure
within a single charging protocol. The spectral tunability and
collective exchange mechanisms considered here are compatible in principle
with weakly anharmonic mediator architectures and provide a route for
controlling how energy and extractable work are distributed in small
quantum batteries.

\bibliographystyle{JHEP}
\bibliography{ref_mod}

\end{document}